# Effect of a fine-scale layered structure of the atmosphere on infrasound signals from fragmenting meteoroids

Igor P. Chunchuzov[1,*], Oleg E. Popov[1], Elizabeth A. Silber[2,3] and Segey N. Kulichkov[1,4]

[1]Obukhov Institute of Atmospheric Physics, Russian Academy of Sciences, Moscow, 119017 Russia
[2]Sandia National Laboratories, 1515 Eubank Blvd. NE, Albuquerque, NM, 87123, US
[3]Department of Earth Sciences, Western University, London, ON, N6A 5B7, Canada
[4]Moscow State University, Moscow, 119991 Russia

[*]E-mail: igor.chunchuzov@gmail.com

**Abstract**

We investigate the influence of a fine-scale (FS) layered structure in the atmosphere on the propagation of infrasound signals generated by fragmenting meteoroids. Using a pseudo-differential parabolic equation (PPE) approach, we model broadband acoustic signals from point sources at altitudes of 35–100 km. The presence of FS fluctuations in the stratosphere (37–45 km) and the lower thermosphere (100–120 km) modifies ray trajectories, causing multiple arrivals and prolonged signal durations at ground stations. In particular, meteoroids fragmenting at 80–100 km can produce two distinct thermospheric arrivals beyond 150 km range, while meteoroids descending to 50 km or below yield weak, long-lived arrivals within the acoustic shadow zone via antiguiding propagation and diffraction. Comparison with observed infrasound data confirms that FS-layered inhomogeneities can account for multi-arrival "N-waves," broadening potential interpretations of meteoroid signals. The results also apply to other atmospheric-entry objects, such as sample return capsules, emphasizing how FS structure impacts shock wave propagation. Our findings advance understanding of wavefield evolution in a layered atmosphere and have broad relevance for global infrasound monitoring of diverse phenomena (e.g., re-entry capsules, rocket launches, and large-scale explosions).

1. **Introduction**

It is recognized now that anisotropic wind velocity and temperature fluctuations caused by internal gravity waves in the atmosphere significantly affect the amplitude, travel time and propagation direction of infrasound waves (Alexis Le Pichon et al., 2019; Chunchuzov and Kulichkov, 2020, Sec.5). This, in turn, leads to errors in determining the location of various infrasound sources (e.g., explosions of different yield, rocket launches, meteorite falls, volcanic eruptions, thunderstorms, hurricanes, tornadoes) and their energy. Correctly accounting for these fluctuations is therefore critical for the infrasound monitoring of natural and anthropogenic events (e.g., Alexis Le Pichon et al., 2019).

In this paper, we investigate the influence of anisotropic wind velocity and temperature fluctuations on the infrasound field of quasi-point sources, such as meteoroids, which generate strong shock waves (often approximated as N-waves) when travelling at hypersonic speeds in the atmosphere (Silber et al.,2023a, Silber, 2024), as well as Sample Return Capsules (Silber et al,, 2023b). We examine how signal fields, calculated by the pseudo-differential parabolic equation (PPE) method (Avilov, 1995; Avilov and Popov, 2018) at various altitudes (100 km, 80 km 50 km, 35 km) compare with recorded signals from meteoroids during their fragmentation episodes, when these sources can be treated as near-point emitters. The objective of such comparison is to explain the observed waveforms, durations, and characteristic periods of infrasound signals as a function of source altitude and distance from the source in the presence of fine-scale layered structure in the real atmosphere.

The main properties of changes in the waveform, duration, and dominant periods of signals as a function of the source height in the real atmosphere with its inherent fine-scale (FS) layered structure have not been systematically studied to date. At the same time, understanding these properties is fundamentally important because the source power at a given altitude is often estimated by the dominant frequency of the signal at the reception point (Silber and Brown, 2014; Silber et al., 2025). This frequency depends not only on the peak of the emitted spectrum near the source, but also on the frequency-dependent transmission properties of the atmosphere in the presence of its FS layering.

The use of the PPE method makes it possible to take into account not only sound refraction and multipathing in large-scale atmospheric inhomogeneities (relative to the wavelength), but also sound scattering and diffraction caused by FS structure – effects that ray theory does not capture.

Recent works by Pilger et al. (2015; 2020; 2021) have modeled the long-range infrasound propagation in atmospheric waveguides using parabolic equation methods for various meteoroids and rocket launches reaching infrasound stations up to ~5000 km away. However, the influence of the FS structure on the waveforms of infrasound signals from meteoroids has not, to our knowledge, been studied as comprehensively as we do here.

For the meteoroids traveling at hypersonic speeds, the rapid compression and heating of the surrounding air generates strong shock waves (often forming an N-wave structure) (Silber et al., 2018). One of the challenges in analyzing infrasound signals from fragmenting meteoroids is distinguishing multiple arrivals caused by multipath propagation through the FS structured atmosphere from those originating at different fragmentation points along the trajectory (Silber, 2014; 2024).

Below, we present for the first time the results of calculations by PPE-method calculations of the acoustic signal field for a point pulse source whose altitude decreases from 100 km to 35 km. In Section 2 we describe the atmospheric models with FS fluctuations of the effective sound speed that are used for PPE calculations of the infrasound field. The effects of atmospheric anisotropic inhomogeneities on infrasound propagation from quasi-point sources as a function of source height are analyzed in Section 3. In this section, we also compare PPE model calculations with observed infrasound signals generated by meteoroid fragmentation at different altitudes.

In Section 4, we consider a simplified atmospheric model with FS structure localized solely in the stratospheric layer to isolate the effect of this layer on the signals traveling through or reflecting from it. The PPE model calculations are also compared in this section with the observations

## 2. Models of anisotropic inhomogeneities of the atmosphere

The vertical profiles of anisotropic fluctuations $\Delta C_{eff}$ ($z,r$) of the effective sound speed (sound speed plus wind velocity projection on sound wave propagation direction) varying as a function of the horizontal distance $r$ from the source were obtained from their formation model developed in (Chunchuzov, 2009; Chunchuzov et al., 2011). These profiles extend from the ground surface ($z$=0) up to the lower thermosphere ($z$=120 km), (Fig.1a,b), and will be used below in the PPE calculations of the infrasound field. By adding these fluctuations (at horizontal steps of 28 km) to an unperturbed effective sound speed profile, we obtain the two-dimensional distribution of the effective sound speed $C_{eff}(z,r)$ along the infrasound propagation path (Fig.1a,b).

For the distribution shown in Fig. 1a,b we used the unperturbed $C_{eff}$ profile obtained in Chunchuzov et al. (2011) from the Ground-2-Space (G2S) atmospheric model (Drob et al., 2008; 2015). The adequacy of this anisotropic inhomogeneity model is supported by the fact that the vertical and horizontal wavenumber spectra obtained from it agree with observed spectra in different stably stratified atmospheric layers, as measured by radars (Tsuda, 2014; Fritts and Alexander, 2003), infrasound sensors (Chunchuzov et al., 2013; Vorobeva et al., 2023) and airplane-based observations (Bacmeister, et al., 1996; Chunchuzov et al., 2024a; 2024b).

To isolate the effect of an FS structure of the stratosphere on the waveform and duration of infrasound signals, we also consider a simplified case in which FS structure is confined solely to the 37-45 km stratospheric layer, remaining constant with increasing horizontal distance $r$ (Fig. 1c,d). The configuration is particularly relevant for ranges less than 300 km, where an acoustic shadow zone can exist, an area into which signals from a downward-moving supersonic source (e.g., a meteoroid shock wave) may penetrate only because of scattering from the FS structure. Such modeling is crucial for understanding how meteor-generated shock waves, which can be approximated as broadband point sources in the upper atmosphere, are transformed as they propagate through layered inhomogeneities, ultimately shaping the recorded infrasound signals at ground-based stations.

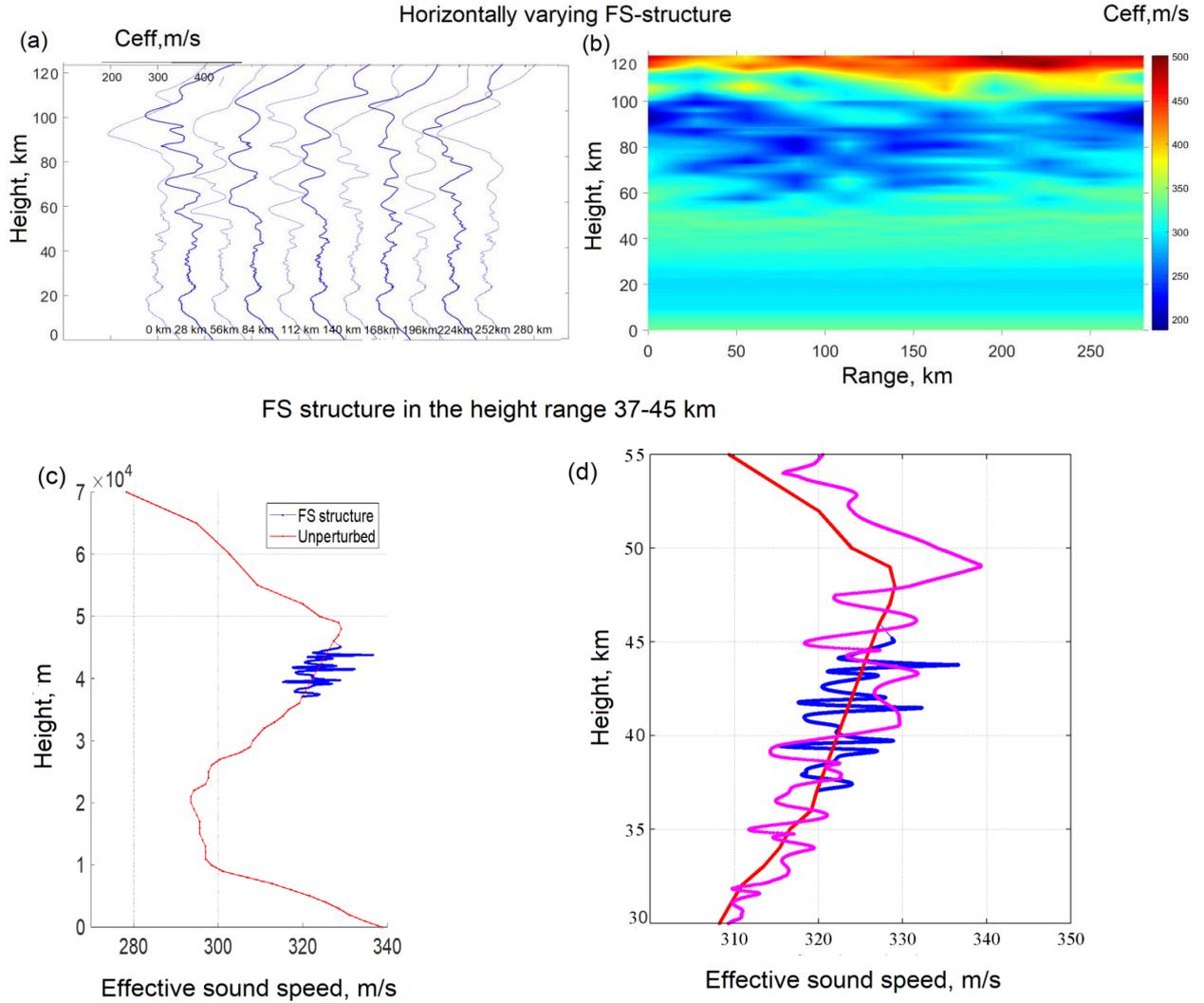

Fig.1. Models of the atmosphere that take into account its fine-scale (FS) layered structure: (a) vertical profiles of the effective sound speed *Ceff* (z) perturbed by anisotropic fluctuations $\Delta Ceff(z,r)$ that depend on the horizontal distance *r* from the source (shown at a horizontal step of 28 km). The unperturbed profile was taken from Chunchuzov et al. (2011), when analyzing infrasound signals from v. Tungurahua (Ecuador); (b) the distribution of the effective sound speed *Ceff(z,r)* along the infrasound propagation path as a function of height *z* and distance *r*; (c) vertical profile of *Ceff* (z) up to 70 km, with added vertical fluctuations caused by FS structure in the 37-45 km stratospheric layer (red line); and (d) its comparison with one of the varying *Ceff(z,r)* profiles (magenta).

## 3. Effect of atmospheric anisotropic inhomogeneities on infrasound propagation from point sources at different altitudes

### 3.1. Sources at altitudes of 90-100 km

Fig. 2 shows the infrasound signals from a point source at an altitude of 100 km, calculated as a function of the horizontal distance *r* from the source (vertical axis) and travel time (horizontal axis). The travel time is referenced relative to the arrival time of sound $t_0=r/c_0$ propagating in the homogeneous atmosphere with some constant sound speed $c_0$=365 m/s taken at an altitude of 100 km. The signals shown in Fig. 2a were calculated using the PPE method in the 0.3-1.5 Hz frequency range, assuming a horizontally varying FS structure (as in Fig. 1a,b). The signal selected near the source, at r=1.72 km (shown below in Fig. 5b) has a uniform frequency spectrum in the frequency range of 0.3–1.5 Hz. Although the original signal is not N-shaped, it allows us to model the "splitting" of the signal into multiple arrivals at large distances from the source.

For comparison, Fig. 2b presents the results in the absence of these fluctuations (i.e. for the unperturbed *Ceff* (*z*) profile). The corresponding ray trajectories for both cases, originating from a point source at 100 km, are shown in Fig. 2c and Fig. 2d, respectively. These trajectories were calculated based on the software package developed by Avilov et al. (2004). The use of ray-trace calculations at mesospheric and lower thermospheric altitudes (60-120 km) is justified because the wavelengths (ranging from hundreds of meters to about 1 km) at 0.3-1.5 Hz are much smaller than the characteristic vertical scales of *Ceff* - inhomogeneities (on the order of tens of kilometers). At these altitudes, meteoroids traveling at hypersonic speeds still generate intense shock waves that often exhibit *N*-wave characteristics, and the short-wavelength signal content is especially sensitive to anisotropic fluctuations in the atmosphere. Consequently, scattering and diffraction caused by fine-scale layered structures can significantly modify the recorded waveforms and arrival times at ground-based sensors.

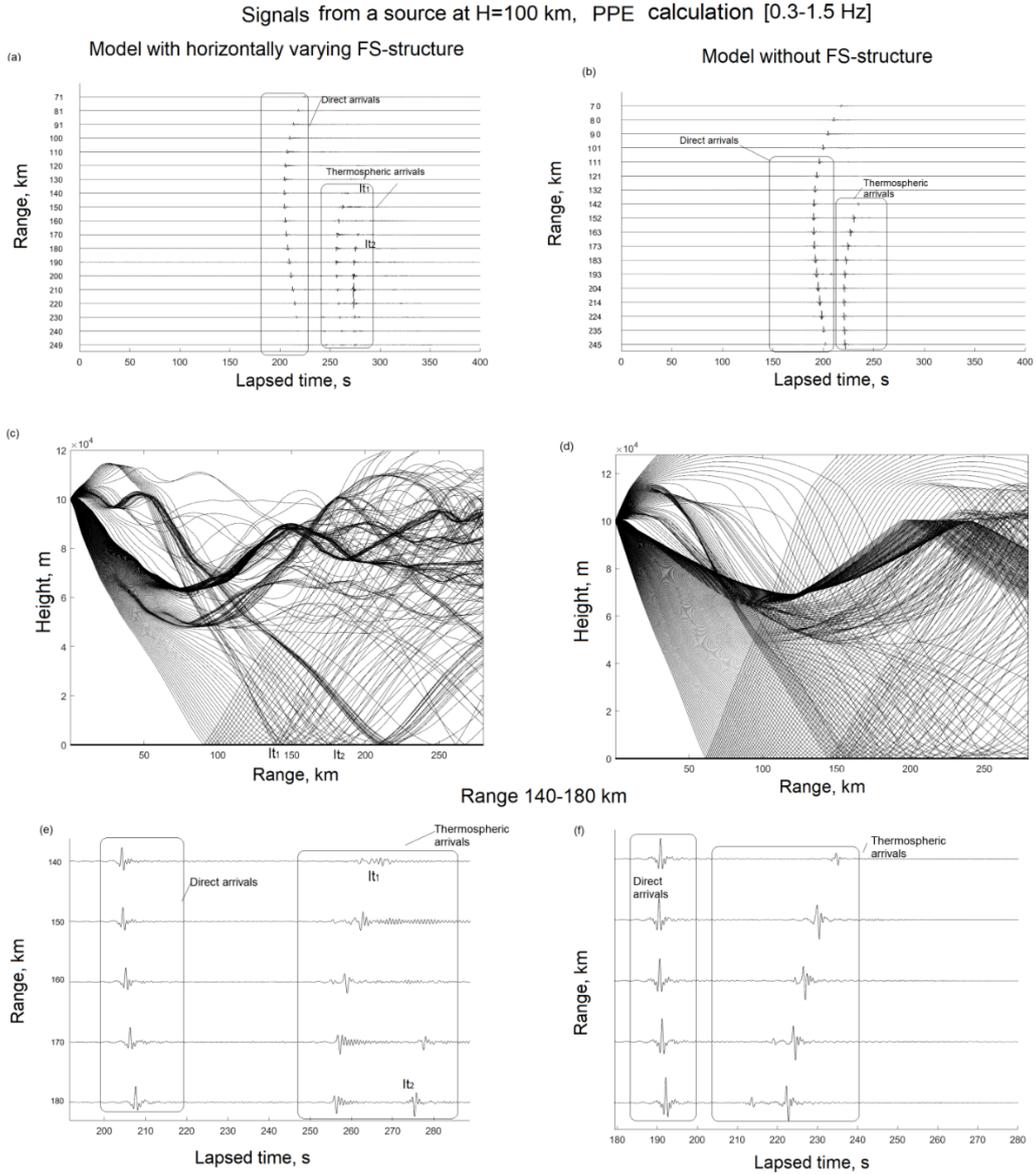

Fig. 2. Signals from the source at an altitude of 100 km, calculated by the PPE method as a function of horizontal distance from the source $r$ (vertical axis), and travel time (horizontal axis). The travel time is measured relative to $t_0 = r/c_0$, where $c_0 = 365$ m/s. Panels (a,c,e) show results for an atmospheric model with a horizontally varying FS structure, while (b,d,f) depict signals for an unperturbed $C_{eff}(z)$ profile. The ray trajectories corresponding to these models are also shown in (c) and (d), respectively.

In both cases, the calculated signals (Fig. 2a and Fig. 2b) show two branches of signal arrivals as a function of the horizontal distance $r$. The first branch (direct arrivals) consists of rays traveling directly from the source to ground-based reception points without reflections. The second branch (thermospheric arrivals) corresponds to rays, labelled $It_1$ and $It_2$, that depart the source at small vertical angles, propagate above the source altitude, and then reflect in the 100-120 km layer of the lower thermosphere before reaching the ground.

The thermospheric arrivals start to appear at a distance of 140 km from the source. However, under the influence of the FS structure, two distinct thermospheric arrivals, $It_1$ and $It_2$, of comparable amplitude emerge at a distance of 170 km (Fig. 2a and Fig. 2e). In contrast, when the FS structure is absent (Fig. 2b and Fig. 2f), a single thermospheric branch is observed.

Notably, FS-induced refraction of rays in the 110-120 km region leads to a multitude of ray paths reflected from the thermosphere, arriving at 170-230 km (Fig.2c), emphasizing the role of fine-scale inhomogeneities in shaping the infrasound signal field at greater distances.

### 3.2. Source at altitude of 80 km

The appearance of two thermospheric arrivals in the presence of varying FS structure, along with the direct arrival, is well illustrated by the ray trajectories from the sources at altitudes of 93 km (Fig. 3a and Fig. 3b) and 80 km (Fig. 3c and Fig. 3d).

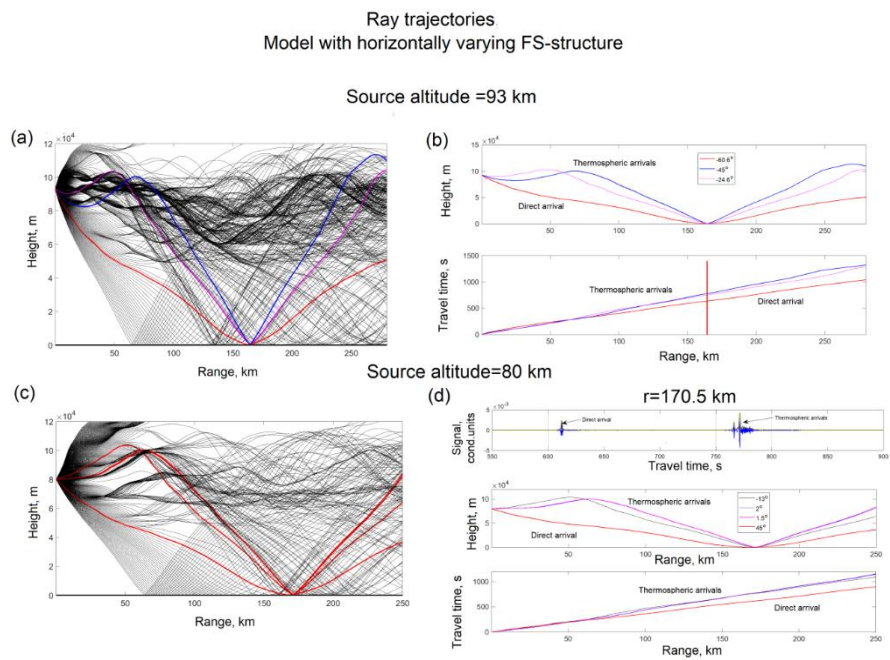

Fig.3. Ray trajectories from the source at altitudes of 93 km (a) and 80 km (c) in an atmospheric model with horizontally varying FS-structure; (b) rays from the source at 93 km altitude and signal travel times for the direct arrival (red) and two reflected arrivals from the lower thermosphere (blue and magenta), arriving at a horizontal distance $r$=164.3 km from the source; (d) the direct and two thermospheric-reflected signal arrivals at $r$=170.5 km and their travel times from the source at an altitude of 80 km. The ray trajectories corresponding to thermospheric-reflected and direct signal arrivals are shown in red (c).

Because the FS-structure includes vertical oscillations of the effective sound speed $C_{eff}(z)$, the ray-path pattern is highly sensitive to the source altitude $H$. This is evident when comparing Fig.3 a,b ($H$=93 km) and Fig.3 c,d ($H$ =80 km). For instance, when the source is at 93 km, it coincides with a local minimum of $C_{eff}(z)$ (see Fig.1a). In this situation, several rays become trapped in local waveguide channels of limited horizontal extent (Fig. 3a). Nevertheless, some rays (thermospheric arrivals) reflect from altitudes near 100 km and reach a horizontal distance of $r$=164.3 km, resulting in two distinct thermospheric arrivals (Fig. 3b).

When the source is at an altitude of 80 km and the FS structure is included, the PPE-based calculations again predict two isolated thermospheric signal arrivals at $r$=170.5 km (Fig. 3d, upper panel). The ray trajectories for these arrivals, along with their travel times, are shown in Fig. 3d (middle and bottom panels). In this case, the onset of two thermospheric arrivals begins at about $r$=160.5 km, as confirmed by the PPE signal calculations in Fig.4a, e, and by the ray trajectories (labeled $I_t$) in Fig. 4c.

In contrast, for the unperturbed atmospheric model (Fig.4b,d,f), the thermospheric branch remains mostly single at distances beyond 160 km, following rays reflected from altitudes around 120 km (Fig.4f). However, at a distance of 180 km, a second, weaker thermospheric arrival can appear before the main arrival; it corresponds to the ray reflected from ~110 km (Fig.4f). Such extra arrivals illustrate how even small vertical gradients in the lower thermosphere can modulate $N$-wave–type shock fronts generated by meteoroids at high altitudes, ultimately creating multiple infrasound arrivals detectable at ground stations.

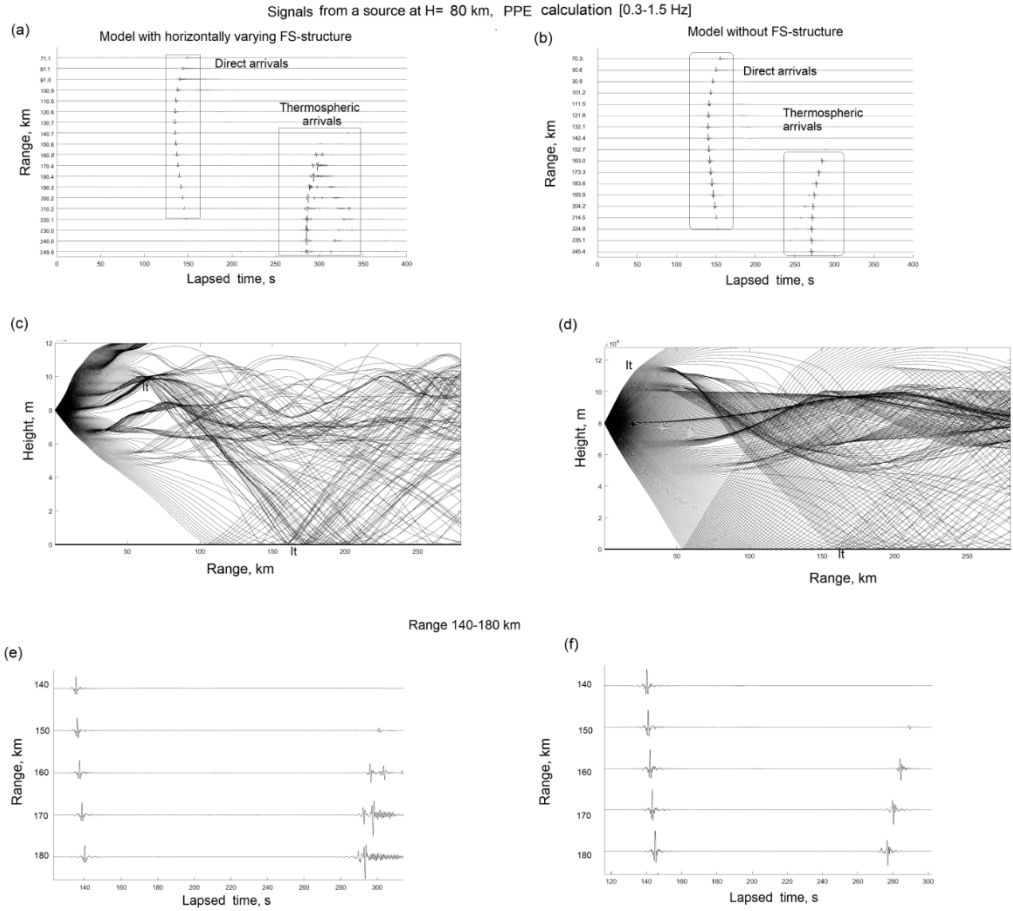

Fig.4. Ray trajectories and signals from a source at 80 km altitude, calculated by the PPE method as a function of horizontal distance $r$ from the source (vertical axis), and travel time (horizontal axis), measured relative to $t_0 = r/c_0$, where $c_0 = 365$ m/s. The branches of direct and thermospheric arrivals are indicated for both perturbed (left panels) and unperturbed (right panels) models.

Fig. 5a shows the 20070421 signal (we use the figures and signal designations from (Silber, 2014)) generated by the fragmentation of a meteoroid at an altitude of 89.2 km, as recorded by four receivers at a horizontal distance of ~172 km. In Fig. 5b, we present the PPE-calculated signals near the source ($r = 1.72$ km), and at $r = 172.3$ km for a source at $H = 89.6$ km, using an atmospheric model with FS structure confined to 37-45 km. As seen, the calculation predicts both the direct arrival and thermospheric arrivals at $r = 172.3$ km.

A correlation analysis, similar to PMCC (Progressive Multi-Channel Correlation) (Cansy and Le Pichon, 2008), was applied to recordings at three receivers, yielding the azimuth and horizontal phase speed of the direct arrival (Fig.5c). The PPE-predicted direct arrival at $r = 172$ km for a source at $H = 89.2$ km with an atmosphere with FS structure aligns well with observations in timing (Fig.5c). Additionally, the model foresees a thermospheric arrival during the time interval 727 - 737 s (Fig.5d, bottom panel). Its passage through the stratospheric layer

The correlation analysis of observed traces also reveals very weak in amplitude arrivals in with FS fluctuations causes the initial meteoroid-generated pulse (Fig.5b, top panel) to split into two distinct arrivals, increasing in the overall thermospheric signal duration by a factor of about three relative to the original radiated signal. 717-733 s (Fig. 5d, upper panel), maintaining an almost constant azimuth (middle panel). However, these arrivals significantly differ in azimuth from the direct arrivals (almost on 180 deg), and are barely distinguishable above the noise, likely due to strong attenuation of high-frequency spectral components (> 1 Hz) in the 100-120 km layer of the lower thermosphere (Sutherland and Bass, 1995). Thus, we doubt that these weak arrivals are those from meteoroid. The frequency-dependent absorption and nonlinear effects were not included in the present model. Nevertheless, the model's prediction of multiple thermospheric arrivals could become more pronounced for more energetic meteoroid events with lower dominant frequencies (<1 Hz), where absorption is reduced and the shock wave retains higher amplitude over greater distances.

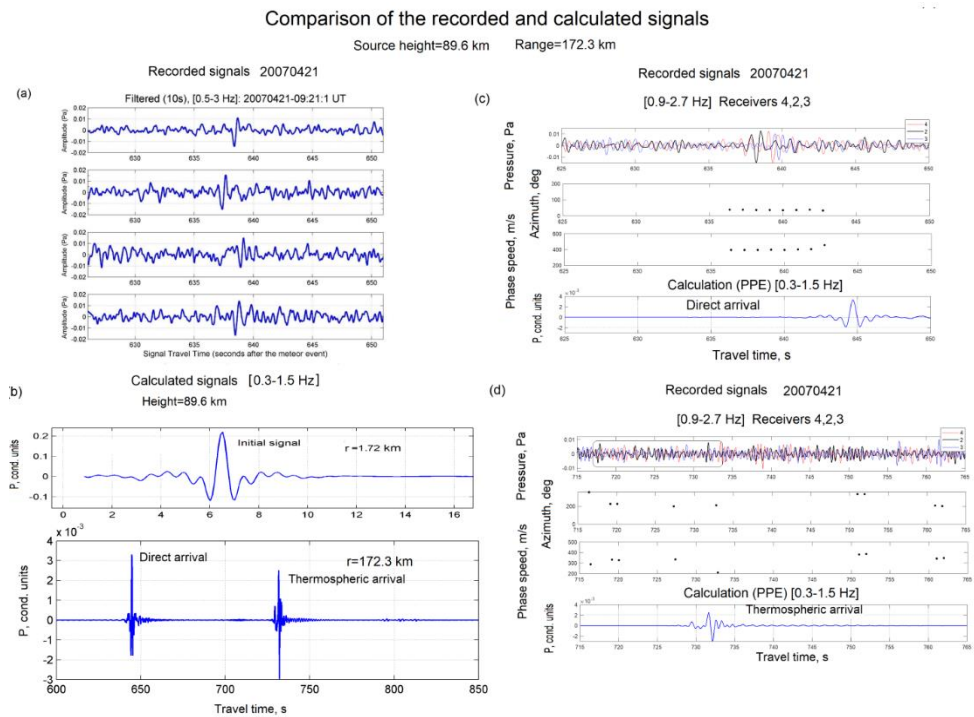

Fig.5. Recorded and PPE-calculated signals at a horizontal distance of $r$=172.3 km from a point source at an altitude of 89.6 km. (a) Signal 20070421, from a meteoroid fragmentation at 89.6 km altitude, recorded by 4 receivers (Silber, 2014). (b) Signals ($P$) for a source at $H$=89.6 km altitude obtained via PPE for an atmospheric model with FS structure at 37-45 km altitudes (see Fig.1c), shown near the source ($r$=1.72 km), and far from it ($r$=172.3 km). (c) Signals recorded by three

receivers in the 625-650 s arrival time range, along with their azimuths and horizontal phase speeds, and the direct signal arrival predicted by PPE. (d) Recorded and calculated signals in the 715-762 s arrival time range.

### *3.3. Sources at altitudes of 50-60 km*

For a source at an altitude of 50 km, Figure 6 compares the calculated signal waveforms at various distances from the source, along with the ray trajectories, under two conditions: one with a horizontally varying FS-structure and one without. In both cases, the calculations predict two primary signal branches: direct arrivals and thermospheric arrivals. The direct arrivals result from downward-propagating rays originating at the source, reaching ground-based receivers at horizontal distances less than ~140 km (Fig. 6, middle panels).

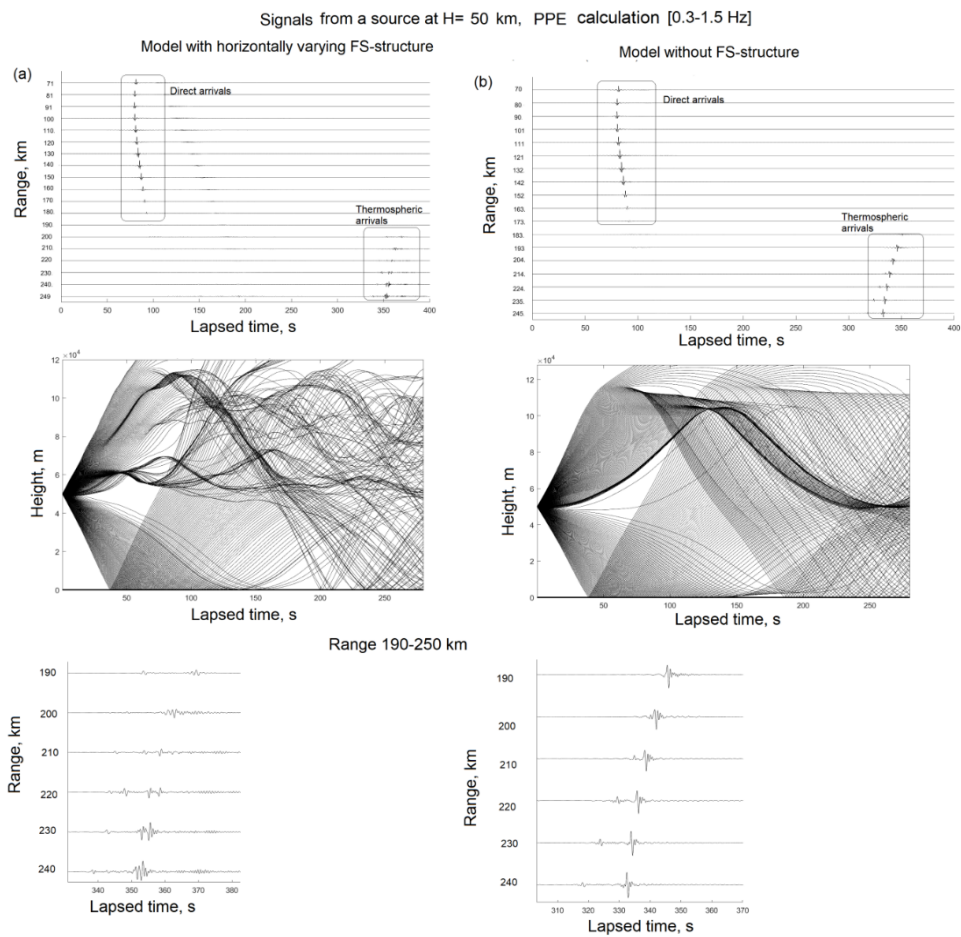

Fig.6. Signal arrivals at different distances from a source located at an altitude of 50 km (PPE calculation), and the corresponding ray trajectories, in an atmospheric model with (a) and without (b) varying FS-structure.

Starting from distances $r \sim 150$ km, rays traveling downward from the source undergo strong refraction up to ray turning point in the surface layer of the atmosphere above ground, where they turn back into the upper atmosphere. Nevertheless, the weak-amplitude arrivals (the direct branch) are predicted even $r = 160$-$170$ km (within the acoustic shadow zone, Fig. 6, upper panel). The appearance of these arrivals in the acoustic shadow zone can be explained by wavefield diffraction, also referred to as antiguiding, where low-frequency signal components penetrate beyond the classical shadow boundary (Brekhovskikh and Lysanov, 1982). Such diffraction effects selectively filter the low-frequency components of the signal spectrum that penetrate in the shadow zone and stretch the pulse duration (Don and Gramond,1986; 1990; Chunchuzov et al., 1990).

Beginning around $r \sim 190$ km, a thermospheric branch appears in both the atmospheric model with FS structure and the unperturbed model (Fig. 6, upper and lower panels). However, in the presence of FS fluctuations, the PPE calculations predict a larger number of thermospheric arrivals and a longer overall signal duration of the thermospheric branch, compared to the case without the FS structure.

## 4. Atmospheric model with FS structure in the stratospheric layer (37-45 km)

### *4.1. Source is above a stratospheric layer (37-45 km)*

To isolate the effect of FS structure in the stratospheric layer on the field of signals propagating from the source to the ground (through the stratopause), we consider a simplified model in which FS fluctuations are included only in the 37-45 km region (Fig. 1c).

In this model, we assume that the FS fluctuations do not vary horizontally up to $r = 300$ km. Compared to an atmosphere with horizontally varying FS (Fig. 1d), the vertical oscillations in $Ceff(z)$ have a shorter characteristic vertical period, while maintaining the same amplitude. This increases the local vertical gradients in the stratospheric layer, allowing us to quantify their effect on the acoustic wave passing through 37–45 km, an important consideration for meteor-generated shock waves that can partially reflect or scatter upon encountering such localized layering.

Fig. 7 shows the PPE-calculated signal waveforms as a function of horizontal range $r$ for a point source at 50 km altitude. The vertical $Ceff(z)$ profile with FS structure in the 37-45 km stratospheric layer is shown in Fig.1c. The corresponding ray trajectories and acoustic field intensity calculated based on software developed in (Avilov, 1995; Avilov et al., 2004; Avilov and Popov, 2018) are shown in Fig. 8a and Fig. 8b, respectively. It can be seen that the trajectories of rays with different elevation angles calculated with step $1^0$ do not fall on the ground surface at distances $r>160$ km (Fig.8a), while the acoustic field penetrates this zone (Fig.8b) due to the effects of diffraction and field scattering (Fig.8c). The calculations predict two branches of arrivals (Fig. 7a). At each distance, waveforms are normalized by their maximum amplitude, which reveals the FS-induced modifications even where amplitudes are strongly attenuated in the acoustic shadow zone ( $r>160$ km, see Fig.8).

As seen in Fig.7b, as $r$ increases from 136 to 167 km, later, weaker arrivals appear after the strongest direct arrival. These later arrivals result from the direct passage of the source signal through the FS-perturbed stratosphere, at angles of incidence that grow with increasing $r$. Moreover, the duration and oscillation period of this first-arriving branch gradually increase with distance from 136 to 167 km, reflecting how fine-scale fluctuations in 37–45 km and subsequent diffraction of the acoustic field in the shadow zone can broaden the pulse and modify its spectral content, a phenomenon especially relevant to the long-range detection of N-wave–type shock fronts from meteoroids or other supersonic and high-energy atmospheric events (e.g., Silber et al., 2023b).

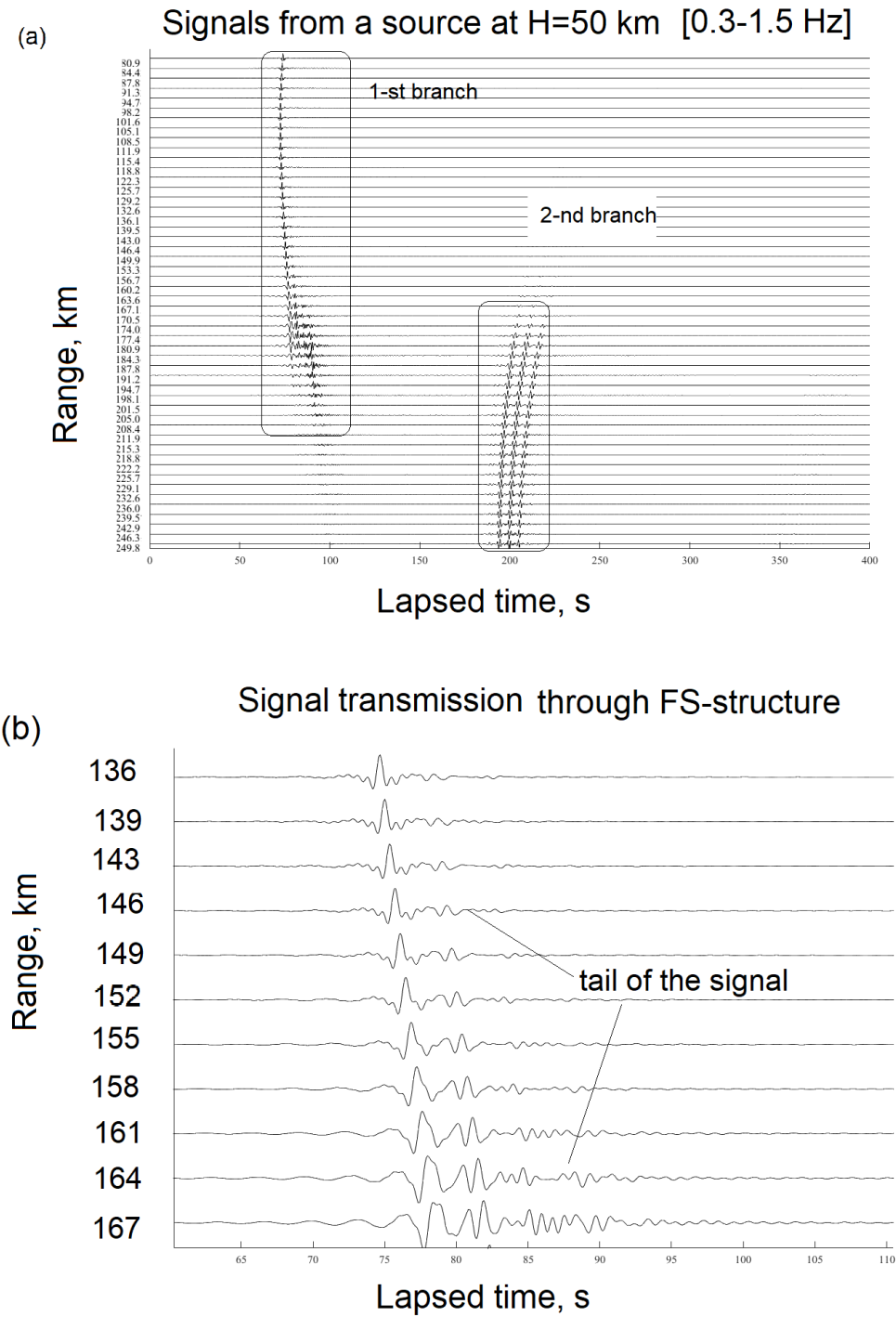

Fig.7. Signals from a source at $H$=50 km calculated via PPE using the vertical $Ceff(z)$ profile with FS structure in the 37-45 km stratospheric layer ( Fig.1c). (a) Direct arrivals (first branch) and arrivals resulting from ground reflections followed by partial reflections from the stratospheric FS layer (second branch). b) Changes in the waveform, oscillation period and overall signal duration as the distance $r$ increases from 136 to 167 km (PPE results).

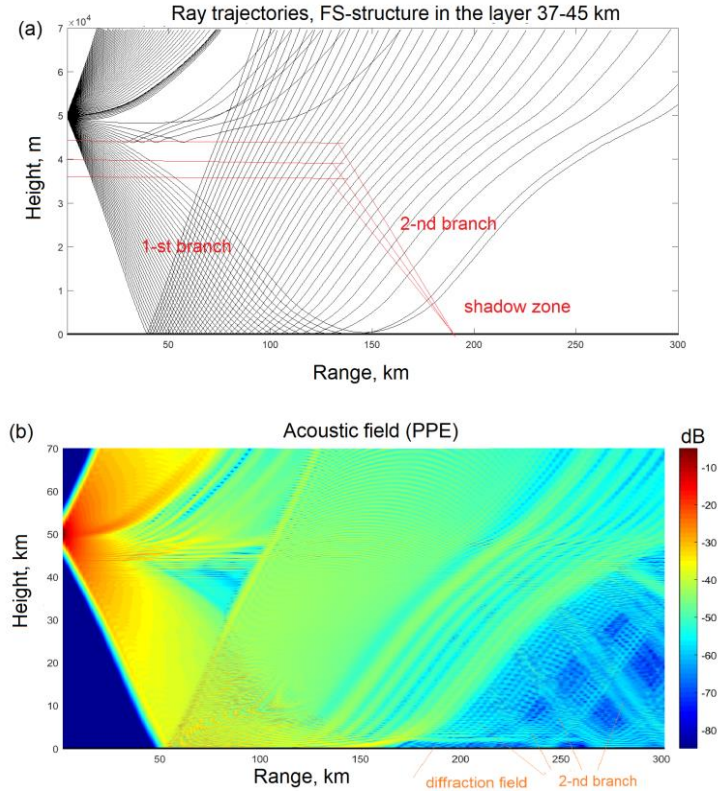

Fig.8. (a) Rays from a source at $H$=50 km for the vertical $C_{\text{eff}}(z)$ profile with a perturbed stratospheric layer (37-45 km), as shown in Fig.1c. (b) Acoustic field intensity distribution (at 1 Hz) from a point source at 50 km altitude.

The appearance of a second arrival branch in the acoustic shadow zone at $r$ >160 km (Fig. 7a) can be attributed to partial reflections from the stratospheric FS layer of waves already reflected from the ground (the first-arrival branch). Fig.8a and Fig.8b illustrate these reflection paths.

### *4.2. Source is below a stratospheric layer (37-45 km)*

In earlier work (Chunchuzov et al., 2013), it was shown that if an *N*-wave impinges at large incidence angles (~60-80 deg) on a layer $0 \leq z \leq H_0$ with FS fluctuations in the effective refractive index $\varepsilon(z)$ (proportional to relative fluctuations in the effective sound speed, see formula (3A)), it is stretched in time into a long wave packet (Fig. 1A, upper panel). This temporal stretching results from multiple partial reflections of the incident wave at each height $z$ within an inhomogeneous layer, where vertical oscillations in the refractive index gradient, $d\varepsilon(z)/dz$, lead to different reflection phases and time lags. At the receiver, these interfering partial reflections summate to produce a lengthened wave packet. Fig. 1A (bottom panel) shows the normalized reflected waveform, P'(t)/P'max, from a FS-fluctuating layer when an *N*-wave is incident, based on both the analytical solution (5A) (Theory) and PPE calculations in the 0.16-1.27 Hz band (Chunchuzov et al., 2013).

Beyond the continuous reflections at each height $z$, there are also transmitted waves that can propagate upward toward the upper boundary $z=H_0$ of the atmospheric layer (see Appendix A). The interference of these transmitted plane waves (each with different frequencies, phases, and angles of incidence) ultimately shapes the observed signal at the reception point.

By employing the PPE method, the numerical calculations account for both the influence of FS fluctuations on the waveform and duration of the transmitted signal and the diffraction of the wavefield into the geometric shadow zone. From the ray trajectories (Fig. 8a) and acoustic field distribution (Fig. 8b), it is evident that a shadow zone forms at $r>160$ km. Nevertheless, finite-amplitude signals are still predicted in this zone, consistent with antiguiding propagation of the low-frequency components (Don and Gramond, 1986; 1990; Chunchuzov et al., 1990). Indeed, as $r$ increases from 152 to 167 km, there is a noticeable increase in the leading-edge duration of the signal (Fig. 7b), illustrating how diffraction and antiguiding propagation broaden the acoustic pulse in the shadow region.

### *4.3. Comparison with observed signals generated by meteoroid fragmentation at different altitudes*

Fig. 9a and Fig. 9b show the 20090926 signal recorded by four receivers, produced by a meteoroid descending from an altitude of 70.8 km ($r=$ 109.1 km) to 20.2 km, ($r=$ 133.2 km) (Silber, 2014). This waveform exhibits a time oscillating "tail" that follows the initial head arrival (which has the maximum amplitude). A correlation analysis of the tail at three receivers reveals that its azimuth and horizontal phase speed are almost identical to those of the head arrival (Fig. 9c). The total signal duration extends to about 10 s, consistent with the long-lasting tails predicted by our model. In particular, the calculations in Fig. 7b for a source at $H=$50 km indicate that such a trailing wave packet can arise from passage of the signal through the FS layer (37–45 km) before reaching the receiver (see Appendix A).

Fig. 9d illustrates one of the calculated signals at $r=$147 km for a source at $H=$50 km, after transmission through the FS fluctuations in 37–45 km. The model reproduces the weak secondary arrivals that appear after the dominant direct arrival, thereby prolonging the overall waveform compared to the original near-source pulse.

Signals can also develop long-duration tails if the source descends below the stratospheric FS layer (37-45 km) and the wave reflects back from those inhomogeneities (Fig. 10). As an example, Fig. 10a shows the meteoroid signal (Silber, 2014) recorded at four receivers for altitudes of 38.9 km (r=107.4 km) and 38.1 km (r=107.7 km). In Fig.10b, we compare PPE-based signals at distances of 100 and 140 km from a source $H$= 35 km for: 1) the simplified stratospheric model with FS fluctuations exclusively in 37–45 km (top panel), and (2) the horizontally varying FS model throughout 0-120 km (middle and bottom panels).

In both atmospheric models, the direct arrival maintains the largest amplitude, but subsequent weaker arrivals interfere to form a prolonged tail (Fig. 10b). These later arrivals arise from partial reflections within the stratospheric layer that sits above the source altitude (see Fig. 1A, Appendix A). As a result, when meteoroids fragment below the stratopause, the layered inhomogeneities can extend and modulate the resulting infrasound wavefield, causing multiple wave arrivals and longer overall signal durations at ground-based receivers.

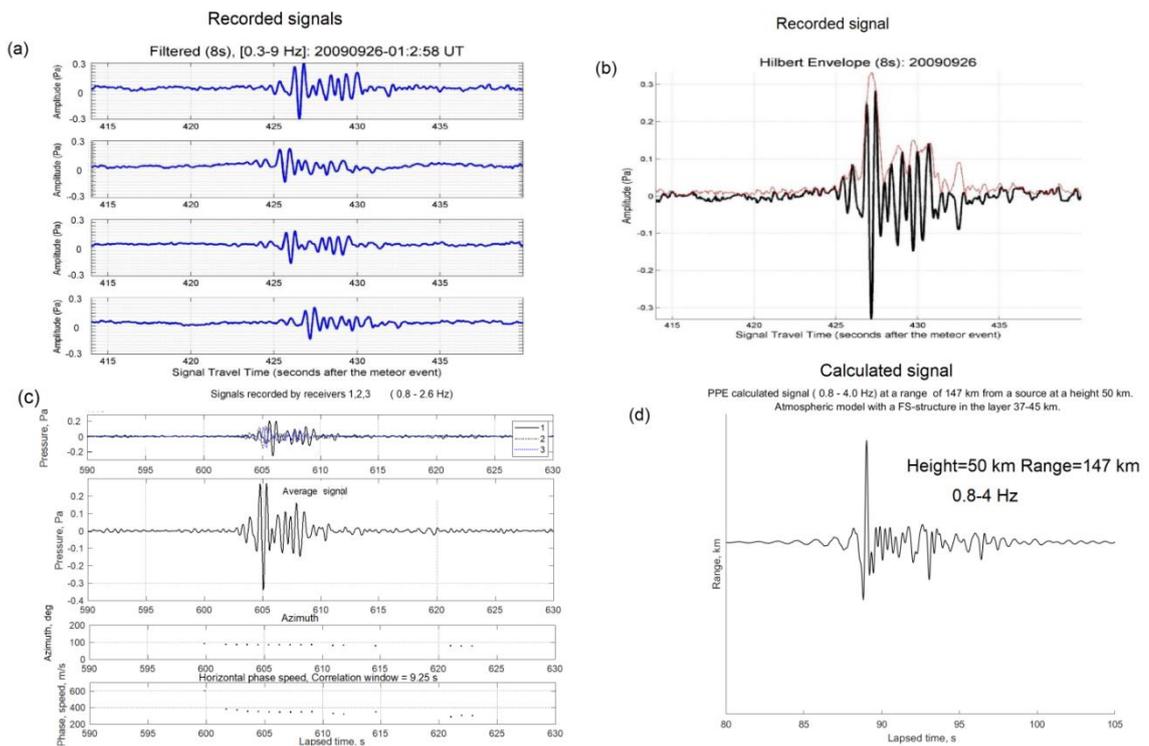

Fig. 9. Signal (20090926) from a meteoroid descending from $H$=70.8 km, ($r$= 109.1 km) to $H$=20.2 km, ($r$= 133.2 km) recorded by four receivers (Silber 2014) (a). The same signal together with its Hilbert envelope. A Hilbert envelope obtained by Hilbert transform of the signal extracts the slowly varying amplitude variations of the signal (or envelope), and removes its rapid oscillations. (b). Average signal obtained after time-delay compensation at receivers 1-2-3, showing azimuth and horizontal phase speed as a function of time (c). Signal waveform from PPE calculations in the 0.8-4 Hz range at $r$=147 km for a source at 50 km altitude.

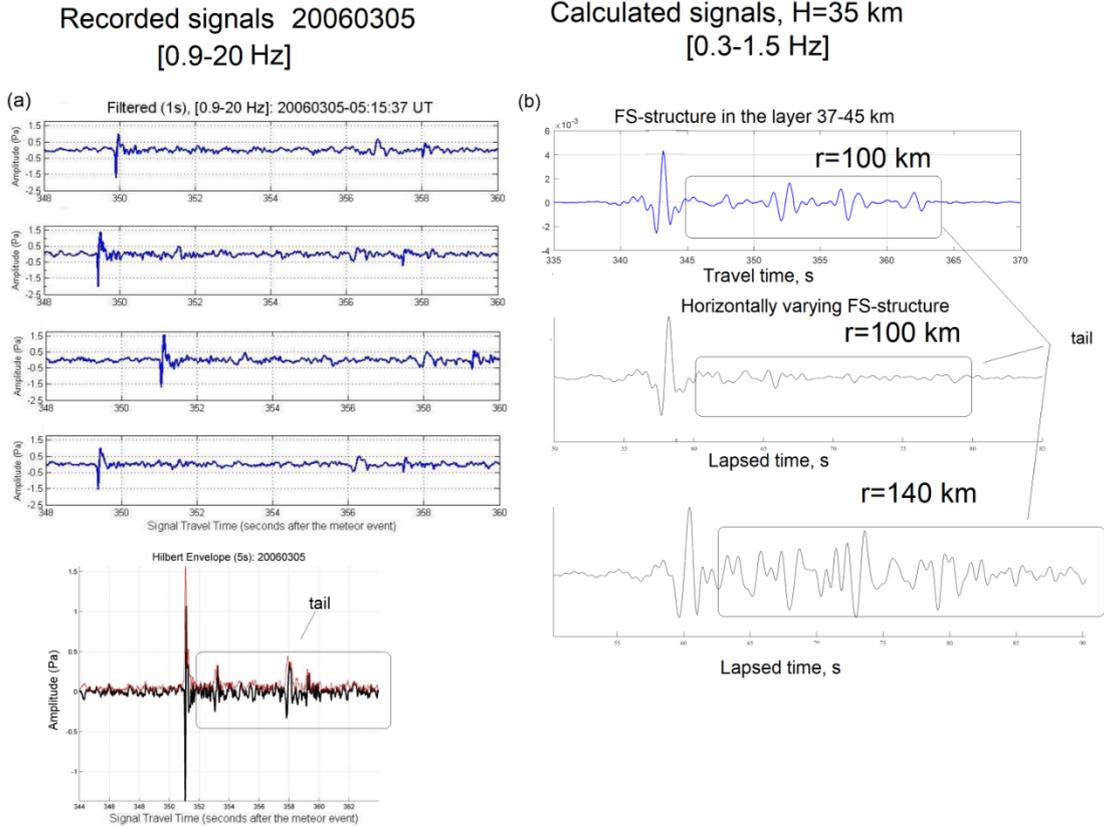

Fig.10 (a) Signal from a meteoroid ($H$=38.1-38.9 km, $r$=107.4-107.7 km) recorded by four receivers (Silber, 2014) and shown with a Hilbert envelope of the signal. (b) Calculated (PPE) signals at 100 km and 140 km from the source at $H$= 35 km, using (i) a model with FS structure restricted to 37-45 km and (ii) a model with horizontally varying FS structure throughout 0-120 km.

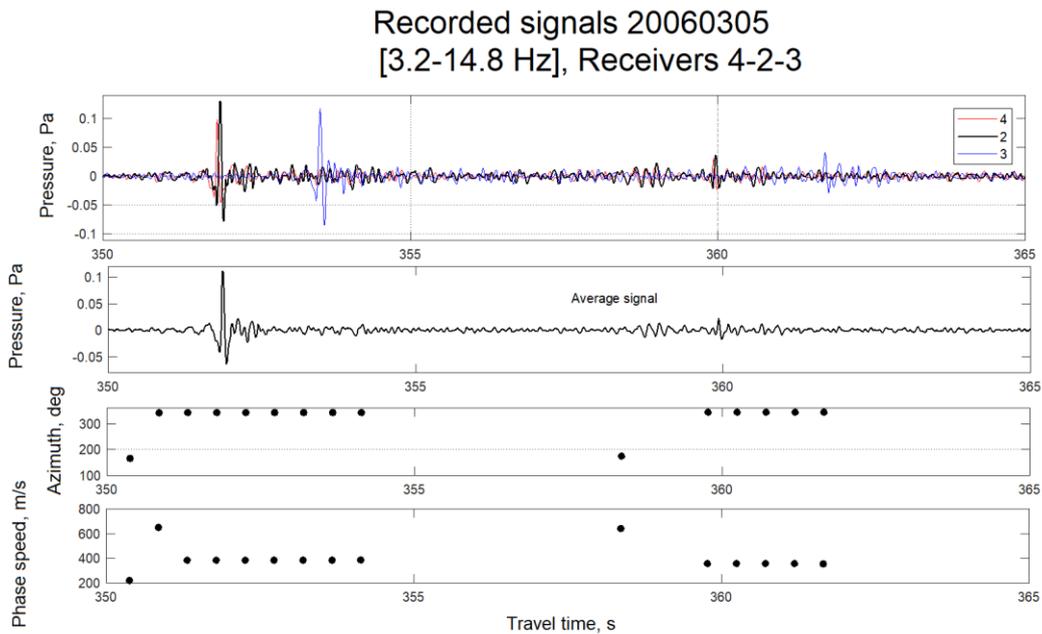

Fig. 11. Average signal obtained after time-delay compensation of the 20060305 signals at receivers 4-2-3, along with the azimuth and horizontal phase speed of the signal as a function of time.

The appearance of the direct arrival (predicted by our model), and the subsequent long-lasting, low-amplitude "tail" are also evident in the recorded signal at ~107 km (Fig. 10a). A correlation analysis of signal traces at three receivers (Fig. 11) confirms that these weak arrivals, which follow immediately after the main pulse (arrival time ~ 352 s), maintain essentially the same azimuth and horizontal phase speed of propagation, with only a short interruption, up to 363 s. Their appearance is explained by partial reflections from the FS fluctuations in the stratosphere, forming a long tail in the calculated signal (see Fig.10b).

The long-lasting oscillating signals from meteoroids fall into the class of "diffuse" multi-arrival signals (Silber, 2014; 2024), attributable to fragmentation events. As shown above, their origin can be tied to passage or reflection within a FS-layered stratosphere, where an initially single arrival is split or lengthened. Similar long-tailed signals have also been detected at the shadow-zone boundary by infrasound sensors on high-altitude balloons ($H$=20 km) at ~169 km from surface explosions (see Fig.2c in (Silber and Bowman, 2023a)).

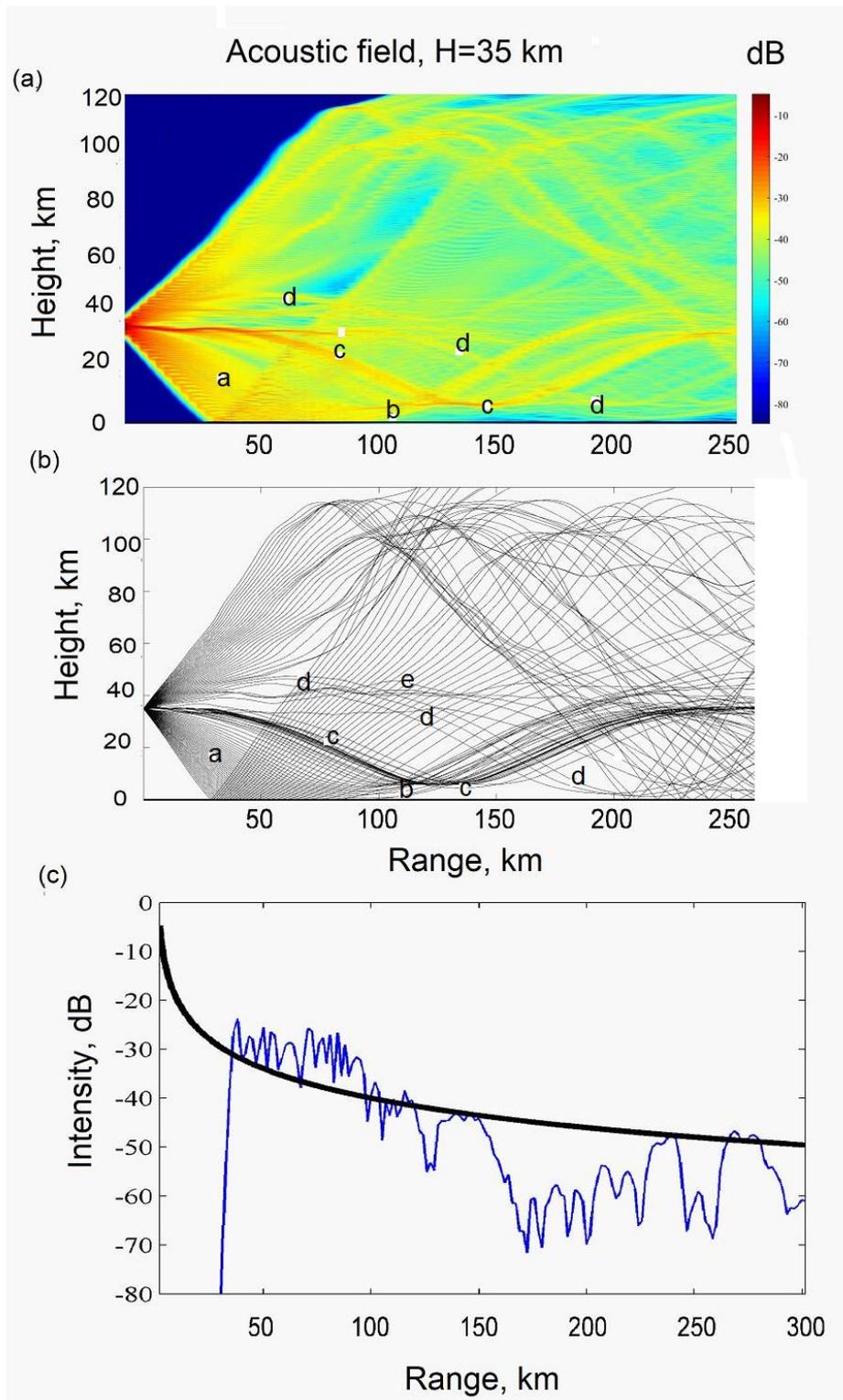

Fig.12. (a) Acoustic field at 1 Hz, (b) ray trajectories from the source at 35 km altitude, and (c) horizontal section of the field intensity at 19 m height (blue), compared with the spherical range $r^{-2}$ dependence from a point source at 19 m (black).

The results of PPE modeling and ray tracing for the source at 35 km altitude are shown in Fig.12a and Fig.12b, respectively, whereas Fig.12c shows horizontal section of the field intensity at 19 m height. Notably, starting at $r$=100 km, the downward-travelling ray paths in Fig.12b labeled a,b undergo reflection from the ground, while c and d have turning points near the ground, turning back upward and thereby forming an acoustic shadow zone at the surface in $100 < r < 220$ km. Nevertheless, effects of diffraction of acoustic field near the focusing points of the ray paths b, c, and d allows weak signals to penetrate into this shadow region. The weak "illumination" of the specified shadow zone at horizontal distances $100 < r < 220$ km can be seen in the acoustic field intensity distribution (Fig. 12a) and in its range dependence at 1.5 Hz (Fig. 12c), where a horizontal cross section is taken at 19 m above the ground.

Thus, signals from $H$= 35 km still appear within the acoustic shadow zone, owing to the combined effect of scattering (or partial reflection) by layered stratospheric inhomogeneities and diffraction near the ground-level caustics. The properties examined here—namely, how source height and horizontal distance influence waveform shape, characteristic periods, and signal durations—must be accounted for when estimating source power at different altitudes and when modeling infrasound from meteoroids in the real, FS-layered atmosphere.

## 5. Conclusion

We have studied the main properties in infrasound waveform changes for descending meteoroids in a real atmosphere containing FS layered structure. The results show that FS fluctuations in the lower thermosphere (100–120 km at 80–) can lead to multipath propagation and generate multiple thermospheric arrivals at ground-based receivers when meteoroids fragment at 100 km altitudes. We considered the fragmenting meteoroids as point sources of N-waves. Similarly, the FS-layered structure affects the N-waves generated from supersonic propagation, but in this case one would have to take into account in the ray tracing and PPE modelling the geometry of the Mach cone and almost cylindrical wave field divergence from meteoroid.

Using atmospheric models perturbed by FS fluctuations of the effective sound speed, we demonstrated that two distinct thermospheric arrivals can appear from the same meteoroid source at horizontal ranges >150 km. These multiple arrivals align with the class of "multi-arrival N-waves produced by fragmentation" (Silber, 2024), yet they can also be misinterpreted as arising from separate fragmentation points if atmospheric multipathing is not accounted for.

For a source at $H$=89.6 km, PPE calculations with an FS-perturbed stratosphere (37-45 km) predicted both direct and thermospheric arrivals at $r$=172.3 km. Observed signals from a meteoroid at a similar altitude show very weak secondary arrivals (their azimuth and phase speed matching those of the direct arrival) although these later arrivals are barely distinguishable above the noise, likely due to strong attenuation of the high frequency (> 1 Hz) content in the 100–120 km layer of the lower thermosphere (not included in our model).

When a meteoroid descends to about 50km before fragmenting, an acoustic shadow zone forms at $r$=150–300 km. At this altitude, downward-propagating rays undergo total reflection in the lower atmosphere, failing to reach the ground and instead turning back upward. Nonetheless, PPE calculations predict that weak-amplitude, long-lasting signals can still reach the shadow zone. This phenomenon arises from two primary effects: (1) the passage of the source-generated wave through the stratospheric FS layer (37–45 km), and (2) antiguiding propagation that carries low-frequency signal components into the shadow region at the surface. The increase of the oscillation period and overall signal duration was predicted by PPE calculations as the distance $r$ increased from 136 to 167 km (Fig.7b). Such an increase in the oscillation period of the signal in the shadow zone can overestimate the energy of its source.

In addition, PPE-based modeling of waves passing through the perturbed stratospheric layer explains the strong direct arrival and the later, weaker arrivals observed in real signals. For a source at 35 km (now beneath the 37–45 km FS layer), the calculations predict long-duration "tails" resulting from partial reflections from the overlying stratospheric layer. A correlation analysis at three receivers confirms the main arrival followed by a lower-amplitude tail, consistent with diffuse multi-arrival signals (Silber, 2014). Notably, FS fluctuations similarly affect *N*-waves, which can radiate nearly cylindrically from a meteoroid.

The main properties identified here, relating waveform characteristics, dominant periods, and signal durations to meteoroid altitude and horizontal range, are essential for accurate modeling of meteor-generated infrasound and for estimating the energy released at different points along the meteoroid's trajectory. Furthermore, these findings are also relevant for the study of other atmospheric-entry objects, such as re-entry capsules (e.g., OSIRIS-REx, Silber et al. (2023b)), which similarly generate shock waves that propagate through the FS-layered atmosphere.

**APPENDIX A**.

The transmission coefficient $W$ (or transparency) of a plane wave through an inhomogeneous layer, incident with wave vector $\mathbf{k}$ at an angle $\theta_0$ on a layer with arbitrary dependence of sound speed and horizontal wind speed on height $z$ (or $C_{eff}(z)$), and passing through this layer, is equal to (Brekhovskikh and Godin, Sec.3):

$$W = (1+V) F^{-1}(z=0) \lim_{z \to \infty} [F(z) \exp(-i\sqrt{(k^2 - k_x^2)} z)], \quad (1A)$$

where $V = V(k_x, z=0)$ is reflection coefficient from this layer depending on the horizontal component of the wave number vector $k_x = k \sin\theta_0$, and $F(z)$ is complex amplitude of acoustic pressure satisfying the Helmholtz equation inside the layer.

For the layer considered here with small fluctuations of the effective refractive index

$$\varepsilon(z) \approx -2\Delta C_{eff}(z)/(c_1 \cos^2\theta) + O(\Delta C_{eff}^2/(c_1 \cos^2\theta)^2), \quad |\varepsilon(z)| \ll 1, \quad (2A)$$

the waveform of the reflected signal for the $N$-wave incident on the layer was obtained in (Chunchuzov et al. 2013) in the single reflection approximation (Fig. 1A, bottom panel).

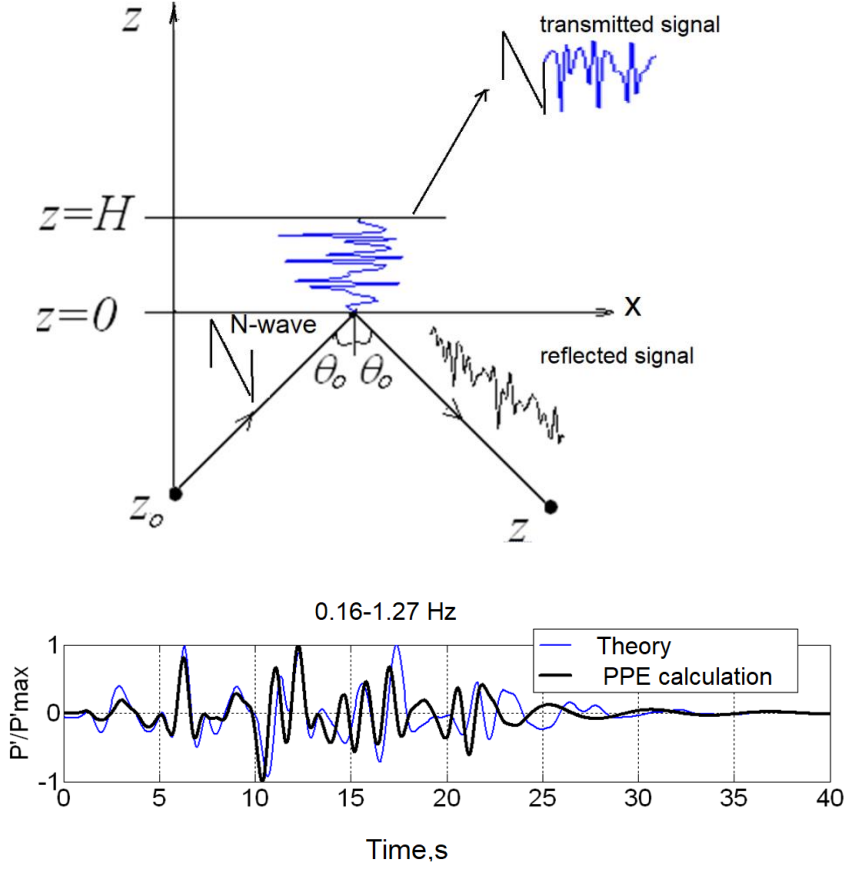

Fig. 1A. Upper panel: Schematic showing incidence of an acoustic N-wave from a point source (located at point $z_0$) on an inhomogeneous layer with vertical fluctuations of the effective refractive index, along with reflected and transmitted signals. The blue portion indicates the signal tail formed as the wave passes through the FS-structured layer. Bottom panel: Waveform of the reflected signal $P'(t)$, normalized to its maximum $P'_{max}$ obtained from both the analytical solution (theory) and PPE calculations in the 0.16-1.27 Hz band (Chunchuzov et al., 2013)

For the spectral component of the incident signal with wave number vector **k**, the contribution to the reflected field in the mirror direction is given by a certain component of the vertical wave number spectrum of the fluctuations $\Delta C_{eff}(z)$ with vertical wave number $K$, satisfying the Bragg condition: $K=2k_z$, where $k_z$ is the vertical component of the incident wave number vector. In this case, the reflection coefficient from the layer is determined by the Fourier spectrum of vertical oscillations of the gradient $d\varepsilon(z)/dz$ of fluctuations of the effective index of refraction:

$$V(k_x, z=0) = i\pi\, k_z\, \tilde{\varepsilon}(2k_z), \qquad (3A)$$

where $\tilde{\varepsilon}(K)$ is a Fourier spectrum of the vertical fluctuations $\varepsilon(z)$, taken in (3A) at

$$K=2k_z = 2\,\omega/c_1\,\cos\theta_0, \qquad (4A)$$

where $\omega$ is the frequency of the spectral component of the pulse, and $c_1$=const is unperturbed sound speed.

The frequency dependence of the absolute value and phase of the transmission coefficient $W$ through the layer (1A), which is proportional to $(1+V)$, is also determined by the vertical wave number spectrum (3A) of the gradient fluctuations $d\varepsilon(z)/dz$. This spectrum determines the waveform and duration of the signal passed through the layer depending on its angle of incidence and, consequently, on the horizontal distance from the source

The Fourier frequency spectrum of the signal reflected from the layer is determined by the product of the frequency spectrum of the incident acoustic pulse and the vertical fluctuation spectrum of the effective refractive index gradient $d\varepsilon(z)/dz$, taken according to Bragg's condition (4) with the vertical wave number of this spectrum equal to $K=2\,\omega/c_1\,\cos\theta_0$. This means that each spectral component of the incident acoustic signal is reflected from the certain component of the vertical fluctuation spectrum of $d\varepsilon(z)/dz$ with vertical wave number $K$, which satisfies Bragg's condition (4A).

In case of $N$-wave incident on the atmospheric layer with fluctuations $\varepsilon(z)$ the wave form of the reflected signal $I_0(t)$ takes on a simple form (Chunchuzov et al., 2013):

$$I_0(t) \approx -[\varepsilon(a(t-R_1/c_1)) + \varepsilon(a(t-R_1/c_1-T_0))]/4. \qquad (5A)$$

where $a \equiv c_1/(2\cos\theta_0)$, $T_0$ is duration of N-wave, and $R_1$ is a total distance from a source to the point of specular reflection from the layer, and from this point to the receiver. This form is shown in Fig.9 (lower panel, theory)

**Acknowledgements:** Sandia National Laboratories is a multi-mission laboratory managed and operated by National Technology and Engineering Solutions of Sandia, LLC (NTESS), a wholly owned subsidiary of Honeywell International Inc., for the U.S. Department of Energy's National Nuclear Security Administration (DOE/NNSA) under contract DE-NA0003525. This article has been authored by an employee of National Technology & Engineering Solutions of Sandia, LLC under Contract No. DE-NA0003525 with the U.S. Department of Energy (DOE). The employee owns all right, title and interest in and to the article and is solely responsible for its contents. The United States Government retains and the publisher, by accepting the article for publication, acknowledges that the United States Government retains a non-exclusive, paid-up, irrevocable, world-wide license to publish or reproduce the published form of this article or allow others to do so, for United States Government purposes. The DOE will provide public access to these results of federally sponsored research in accordance with the DOE Public Access Plan https://www.energy.gov/downloads/doe-public-access-plan. This paper describes objective technical results and analysis. Any subjective views or opinions that might be expressed in the paper do not necessarily represent the views of the U.S. Department of Energy or the United States Government.

This work was partially supported by the Ministry of Science and Higher Education of the Russian Federation: project FMWR-2025-0005 (Sections 3-4).

**Author Contribution** All authors took part in processing and interpretation of the meteoroid generated infrasound data and reviewing the manuscript.

**Conflict of interest:** The authors declare no conflict of interest.